\documentclass[pre,reprint,pdftex]{revtex4-1}
\usepackage{amsmath,amssymb,amsthm}
\usepackage[pdftex,colorlinks]{hyperref}

\newcommand{\bra}[1]{\left\langle #1 \right|}
\newcommand{\ket}[1]{\left| #1 \right\rangle}

\newcommand{\noisecorrelation}[0]{\alpha}
\newcommand{\noisecoefficient}[0]{A}
\begin{document}

\title{Non-Equilibrium Fluctuation-Dissipation Inequality and \\ Non-Equilibrium Uncertainty Principle}
\author{C. H. Fleming and B. L. Hu}
\affiliation{Joint Quantum Institute and Department of Physics, University of Maryland, College Park, Maryland 20742}
\author{Albert Roura}
\affiliation{Max-Planck-Institut f\"ur Gravitationsphysik (Albert-Einstein-Institut), Am M\"uhlenberg 1, 14476 Golm, Germany}
\date{\today}

\begin{abstract}
The \emph{fluctuation-dissipation relation} is usually formulated for a system interacting with a heat bath at finite temperature in the context of linear response theory, where only small deviations from the mean are considered.
We show that for an open quantum system interacting with a non-equilibrium environment, where temperature is no longer a valid notion, a \emph{fluctuation-dissipation inequality} exists.
Clearly stated, quantum fluctuations are bounded below by quantum dissipation, whereas classically the fluctuations can be made to vanish.
The lower bound of this inequality is exactly satisfied by (zero-temperature) quantum noise and is in accord with the Heisenberg uncertainty principle,
both in its microscopic origins and its influence upon systems.
Moreover, it is shown that the non-equilibrium fluctuation-dissipation relation determines the non-equilibrium uncertainty relation in the weak-damping limit.
\end{abstract}

\maketitle

\section{Open Quantum System}
The \emph{fluctuation-dissipation relation} (FDR) is usually formulated for a system interacting with a heat bath at finite temperature in the context of linear response theory, where only small deviations from the mean are considered.
We show that for an open quantum system interacting with a non-equilibrium environment, where temperature is no longer a valid notion, a \emph{fluctuation-dissipation inequality} (FDI) exists.
Clearly stated, quantum fluctuations are bounded below by quantum dissipation, whereas classically the fluctuations can be made to vanish.
The lower bound of this inequality is exactly satisfied by (zero-temperature) quantum noise and is in accord with the Heisenberg uncertainty principle (HUP).
FDI violating quantum noise can be viewed as arising from HUP violating states of the environment and can induce HUP violating states in the open system.
In fact, in the weak-damping limit the non-equilibrium FDR (which must satisfy the FDI) precisely determines the non-equilibrium uncertainty relation (which must satisfy the HUP).

In the following section we present the necessary background material of quantum open systems,
wherein we formally categorize quantum noise in terms of its time dependence, dissipation, and microscopic origin.
Readers familiar with this may skip to our results in the latter sections and refer back as needed.
In the penultimate section we derive the FDI from a microscopic model and contrast it to the usual thermal FDR.
In the final section we work from the other end and motivate the FDI phenomenologically, but less generally.
This result also produces the non-equilibrium uncertainty relation for quantum Brownian motion, which can be contrasted to the finite-temperature uncertainty relation \cite{HuZhang93,HuZhang95,Anderson93,Anastopoulos95}.

\section{Noise and Dissipation} \label{sec:Noise}

\subsection{Open Systems and Noise}
Consider a quantum system weakly interacting with an environment with interaction Hamiltonian:
\begin{equation}
\mathbf{H}_\mathrm{I} = \sum_n \mathbf{L}_n \otimes \mathbf{l}_n \, , \label{eq:Hint}
\end{equation}
expanded as a sum of separable operators, where $\mathbf{L}_n$ and $\mathbf{l}_n$ are system and environment operators respectively. The environment coupling operators $\mathbf{l}_n$ will typically be collective observables of the environment, with dependence upon very many modes.
The system-environment interaction will be treated perturbatively and so the central ingredient is the (multivariate) correlation function of the environment:
\begin{equation}
\noisecorrelation_{nm}(t,\tau) = \left\langle \underline{\mathbf{l}}_n\!(t) \, \underline{\mathbf{l}}_m\!(\tau) \right\rangle_\mathrm{E} \, , \label{eq:alpha}
\end{equation}
where $\underline{\mathbf{l}}_n\!(t)$ represents the time-evolving $\mathbf{l}_n$ in the interaction (Dirac) picture.
In the \emph{influence functional} formalism \cite{Feynman63} for the quantum Brownian model with bilinear couplings between the system and its environment \cite{CaldeiraLeggett83,HPZ92,HPZ93} the correlation function appears as the kernel in the exponent of a Gaussian influence functional, 
called the influence kernel $\zeta$ in Refs.~\cite{Raval96,Raval97}.
Alternatively, in \emph{quantum state diffusion} \cite{Strunz96} this kernel takes the explicit role of a noise correlation for complex Gaussian noise.
The influence kernel, or equivalently, the complex correlation function, can be written as a sum of two real parts corresponding to the noise and dissipation kernels \cite{Raval96,Raval97}:
\begin{equation}
\underbrace{\boldsymbol{\noisecorrelation}(t,\tau)}_\mathrm{complex \, noise} = \underbrace{\boldsymbol{\nu}(t,\tau)}_\mathrm{noise} + \, \imath\! \underbrace{\boldsymbol{\mu}(t,\tau)}_\mathrm{dissipation} \, . \label{eq:3kernels}
\end{equation}
The noise kernel $\boldsymbol{\nu}$ appears in the influence kernel as the 
correlation of an ordinary real stochastic source,
whereas the dissipation kernel $\boldsymbol{\mu}$ alone would produce a purely homogeneous (though not positivity preserving in general) evolution.
These same roles can also be inferred from the Heisenberg equations of motion for the system operators after integrating the environment dynamics,
producing the so-called \emph{quantum Langevin equation} \cite{FordOconnell88}.
Generalized to nonlinear systems and environments (yet still Gaussian in the influence functional), the environment-integrated Heisenberg equations of motion can be expressed
\begin{eqnarray}
\dot{\mathbf{S}}(t) &=& +\imath [ \mathbf{H} , \mathbf{S}(t) ] +\imath \sum_n \left[ \mathbf{L}_n(t) , \mathbf{S}(t) \right] \mathbf{l}_n(t) \, , \\
\mathbf{l}_n(t) &\equiv& \boldsymbol{\xi}_n(t) - 2 \sum_m \int_0^t \!\! d\tau \, \mu_{nm}(t,\tau) \, \mathbf{L}_m(\tau) \, ,
\end{eqnarray}
for the equation of motion of any system operator $\mathbf{S}(t)$,
where $\boldsymbol{\xi}_n(t)$ is an operator-valued stochastic process with symmetrized two-time correlation $\nu_{nm}(t,\tau)$ and commutator $2\imath \mu_{nm}(t,\tau)$.
The commutator is typically deterministic for quantum Brownian motion.

Using the notation of Ref.~\cite{QOS}, the second-order master equation \cite{Kampen97,Breuer01,Strunz04}
of the reduced density matrix $\boldsymbol{\rho}$ can be represented in terms of the noise correlation as
\begin{equation}
\dot{\boldsymbol{\rho}} = \left[ -\imath \mathbf{H}, \boldsymbol{\rho} \right] + \boldsymbol{\mathcal{L}}_2 \{ \boldsymbol{\rho} \} \, ,
\end{equation}
with the second-order contribution given by the operation
\begin{equation}
\boldsymbol{\mathcal{L}}_2 \{ \boldsymbol{\rho} \} \equiv \sum_{nm} \left[ \mathbf{L}_n, \boldsymbol{\rho} \, (\mathbf{\noisecoefficient}_{nm}\! \diamond \mathbf{L}_m)^\dagger - (\mathbf{\noisecoefficient}_{nm}\! \diamond \mathbf{L}_m) \, \boldsymbol{\rho} \right] \, , \label{eq:WCGME}
\end{equation}
where the $\mathbf{\noisecoefficient}$ operators and $\diamond$ product define the second-order operators
\begin{equation}
(\mathbf{\noisecoefficient}_{nm}\! \diamond \mathbf{L}_m)(t) \equiv \int_0^t \!\! d\tau \, \noisecorrelation_{nm}(t,\tau) \, \left\{ \mathbf{G}_0(t,\tau) \, \mathbf{L}_m(\tau) \right\} \, , \label{eq:WCOG}
\end{equation}
given the free system propagator $\mathbf{G}_0(t,\tau): \boldsymbol{\rho}(\tau) \to \boldsymbol{\rho}(t)$.
One context in which the influence functional, Langevin equation, and master equation all work together seamlessly is in the quantum Brownian motion of linear systems \cite{CRV03,QBM}.
In addition to a a quantum Langevin equation for non-commuting operators, linearity makes it possible in that case to have a Langevin equation for real classical stochastic processes from which general quantum correlation functions and the master equation can be exactly derived \cite{QBM}.

\subsection{The Quantum Noise Correlation}
From its microscopic definition, Eq.~\eqref{eq:alpha}, the environmental correlation function is Hermitian in the sense of
\begin{equation}
\boldsymbol{\noisecorrelation}(t,\tau) = \boldsymbol{\noisecorrelation}^{\!\dagger}\!(\tau,t) \, , \label{eq:alpha_her}
\end{equation}
and also positive definite in the sense of
\begin{equation}
\int_0^t \!\! d\tau_1 \! \int_0^t \!\! d\tau_2 \, \mathbf{f}^\dagger(\tau_1) \, \boldsymbol{\noisecorrelation}(\tau_1,\tau_2) \, \mathbf{f}(\tau_2) \geq 0 \, , \label{eq:posdef1}
\end{equation}
for all vector functions $\mathbf{f}(t)$ indexed by the noise.
Positivity and the noise decomposition \eqref{eq:3kernels} are the key properties from which the FDI arises.
In next section we proceed to categorize the relevant time-dependence and corresponding microscopic origins of the environmental correlations.
Then we will relate these features to the categorization of environments into resistive, amplifying, and indefinite.

\subsection{Categorization of Noise} \label{sec:stationary}
\emph{Stationary} correlations are defined by their invariance under time
translations,
\begin{equation}
\boldsymbol{\noisecorrelation}(t,\tau) = \boldsymbol{\noisecorrelation}(t-\tau) \, ,
\end{equation}
and can produce asymptotically stationary (time-constant) master equations.
Such correlations are produced when the environment is in an initially stationary state and its coupling operators in the Schr\"odinger picture are constant in time.
\begin{equation}
\boldsymbol{\rho}_\mathrm{E}(0) = \sum_i p_\mathrm{E}\!\left(\varepsilon_i\right) \ket{\varepsilon_i} \!\! \bra{ \varepsilon_i } \, ,
\end{equation}
yielding the correlation function
\begin{equation}
\noisecorrelation_{nm}(t,\tau) = \sum_{ij} p_\mathrm{E}\!\left(\varepsilon_i\right) \bra{ \varepsilon_i }  \mathbf{l}_n \ket{ \varepsilon_{ij} } \overline{ \bra{ \varepsilon_i }  \mathbf{l}_m \ket{ \varepsilon_{ij} } } \, e^{+ \imath \varepsilon_j (t-\tau)} \, , \label{eq:alpha_stat}
\end{equation}
where $\varepsilon_{ij} \equiv \varepsilon_i - \varepsilon_j$, $\ket{ \varepsilon }$ denotes the energy basis of the environment and $p_\mathrm{E}\!\left(\varepsilon\right)$ are its stationary probabilities at the initial time.
The accompanying characteristic function can be obtained quite directly from the mode sum
\begin{equation}
\boldsymbol{\noisecorrelation}(t) = \frac{1}{2\pi} \int_{-\infty}^{+\infty} \! d\omega \, e^{+\imath \omega t} \, \tilde{\boldsymbol{\noisecorrelation}}(\omega) \, ,
\end{equation}
yielding the Fourier transform
\begin{equation}
\tilde{\noisecorrelation}_{nm}(\omega) \,\underline{\propto}\, 2 \pi \sum_i p_\mathrm{E}\!\left(\varepsilon_i\right) \bra{ \varepsilon_i }  \mathbf{l}_n \ket{ \varepsilon_{i}-\omega } \overline{ \bra{ \varepsilon_i }  \mathbf{l}_m \ket{ \varepsilon_{i}-\omega } } \, , \label{eq:ideal_bath}
\end{equation}
where the underscored proportionality here is strictly in reference to the continuum limit of the reservoir which relates environmental mode sums to integrals given the infinitesimal strength of individual environmental mode couplings.
This can be more rigorously defined through the use of a finite spectral density function in place of the infinitesimal environment couplings.

Also of note are \emph{quasi-stationary} correlations of the form
\begin{equation}
\boldsymbol{\noisecorrelation}(t,\tau) = \boldsymbol{\noisecorrelation}_\mathrm{S}(t-\tau) + \boldsymbol{\delta \noisecorrelation}(t+\tau) \, ,
\end{equation}
where $\boldsymbol{\noisecorrelation}_\mathrm{S}(t-\tau)$ denotes a stationary correlation function,
or, more specifically, the stationary projection of $\boldsymbol{\noisecorrelation}(t,\tau)$,
while $\boldsymbol{\delta\noisecorrelation}(t+\tau)$ is an additional non-stationary contribution.
Such correlations will result from (constant) linear coupling to an environment with non-stationary initial state, such as a squeezed thermal state.
In these cases the stationary projection of the correlation function does correspond to the stationary projection of the initial state of the environment.
As for the non-stationary contributions, due to their highly oscillatory behavior in the late-time limit
they typically lose effect asymptotically.
Therefore, quasi-stationary correlations can produce an asymptotically stationary master equation with equivalent asymptotics as generated by their corresponding stationary correlation.

Finally one can also consider \emph{cyclo-stationary} and \emph{quasi-cyclo-stationary} correlations,
which exhibit a periodic time-translation invariance
and can produce asymptotically cyclo-stationary master equations.
Much of this work can be easily generalized to these cases,
with the exception of detailed balance which cannot be maintained.

\subsection{Noise Decomposition}
\label{sec:NoiseDecomp}
Motivated by the influence functional formalism and Heisenberg equations of motion,
correlation functions of second order can always be decomposed into a real noise kernel and dissipation kernel as in Eq.~\eqref{eq:3kernels}:
The Hermiticity stated in Eq.~\eqref{eq:alpha_her} leads to the relations
\begin{eqnarray}
\boldsymbol{\nu}(t,\tau) & \equiv & \frac{1}{2} \left[ \boldsymbol{\noisecorrelation}(t,\tau) + \boldsymbol{\noisecorrelation}^{\!\mathrm{T}\!}(\tau,t) \right] \label{eq:nuKernel} \, , \\
\boldsymbol{\mu}(t,\tau) & \equiv & \frac{1}{2\imath} \left[ \boldsymbol{\noisecorrelation}(t,\tau) - \boldsymbol{\noisecorrelation}^{\!\mathrm{T}\!}(\tau,t) \right] \label{eq:muKernel} \, .
\end{eqnarray}
The two kernels naturally decompose the second-order operators of Eq.~\eqref{eq:WCOG} into their Hermitian and anti-Hermitian parts, in the ordinary sense of Hilbert space operators:
\begin{eqnarray}
\mathbf{\noisecoefficient}_{nm} &=& \underbrace{\mathbf{N}_{nm}}_\mathrm{diffusion} + \imath\!\!\! \underbrace{\mathbf{M}_{nm}}_\mathrm{dissipation} \, , \\
(\mathbf{N}_{nm} \diamond \mathbf{L}_m) & \equiv & \int_0^t \!\! d\tau \, \nu_{nm}(t,\tau) \, \left\{ \mathbf{G}_0(t,\tau) \, \mathbf{L}_m(\tau) \right\} \, , \\
(\mathbf{M}_{nm} \diamond \mathbf{L}_m) & \equiv & \int_0^t \!\! d\tau \, \mu_{nm}(t,\tau) \, \left\{ \mathbf{G}_0(t,\tau) \, \mathbf{L}_m(\tau) \right\} \, ,
\end{eqnarray}
The second-order master equation $\boldsymbol{\mathcal{L}}_2 \{ \boldsymbol{\rho} \}$ can then be expressed entirely in terms of Hermitian operators as
\begin{equation}
- \sum_{nm} \left[ \mathbf{L}_n , \imath \left\{ (\mathbf{M}_{nm} \diamond \mathbf{L}_m) , \boldsymbol{\rho} \right\} +  \left[ (\mathbf{N}_{nm} \diamond \mathbf{L}_m) , \boldsymbol{\rho} \right] \right] \, . \label{eq:GHPZ}
\end{equation}
Here the noise coefficients describe diffusion while the dissipation coefficients describe dissipation (or amplification), renormalization and other homogeneous dynamics.

The correlation function $\boldsymbol{\noisecorrelation}(t,\tau)$ is positive definite and, therefore, the noise kernel $\boldsymbol{\nu}(t,\tau)$ must also be positive definite.
The dissipation kernel $\boldsymbol{\mu}(t,\tau)$ is not positive definite, but it is related to the damping kernel $\boldsymbol{\gamma}(t,\tau)$, which is given by
\begin{equation}
\boldsymbol{\mu}(t,\tau) = -\frac{\partial}{\partial \tau} \boldsymbol{\gamma}(t,\tau) \, , \label{eq:damping_kernel}
\end{equation}
and can be positive definite, negative definite, or indefinite.
The dissipation kernel coefficients can then be expressed in terms of damping kernel coefficients
\begin{eqnarray}
\underbrace{(\mathbf{M}_{nm} \diamond \mathbf{L}_m)}_{\mathrm{dissipation}} &=& \underbrace{(\boldsymbol{\Gamma}_{nm} \diamond \dot{\mathbf{L}}_m)(t)}_{\mathrm{damping}} - \underbrace{\gamma_{nm}(t,t) \, \mathbf{L}_m(t)}_{\mathrm{renormalization}} \nonumber \\
&&+ \underbrace{\gamma_{nm}(t,0) \left\{ \mathbf{G}_0(t) \, \mathbf{L}_m(0) \right\}}_{\mathrm{slip}} \, \\
(\boldsymbol{\Gamma}_{nm} \diamond \dot{\mathbf{L}}_m) & \equiv & \int_0^t \!\! d\tau \, \gamma_{nm}(t,\tau) \, \left\{ \mathbf{G}_0(t,\tau) \, \dot{\mathbf{L}}_m(\tau) \right\} \, , \\
\dot{\mathbf{L}}_m(t) & \equiv & +\imath\left[ \mathbf{H}(t) , \mathbf{L}_m(t) \right] + \frac{\partial}{\partial t} \mathbf{L}_m(t) \, , \label{eq:Ldot}
\end{eqnarray}
which can be used to place Eq.~\eqref{eq:GHPZ} into a form much like the QBM master equation \cite{HPZ92,QBM,NQBM}.
The slip is a transient effect as a result of dealing with a factorized initial state, which is modified if one considers correlated initial states.
The renormalization is a permanent shift of the system Hamiltonian which would diverge in the limit of local damping and is typically canceled with a counterterm introduced with the bare system action or the system-environment interaction.

For stationary correlations $\boldsymbol{\noisecorrelation}(t-\tau)$ with characteristic function (Fourier transform) $\tilde{\boldsymbol{\noisecorrelation}}(\omega) = \int_{-\infty}^{+\infty} d\tau \, e^{-\imath \omega \tau} \boldsymbol{\noisecorrelation}(\tau)$, the noise and damping kernels are then Hermitian in both noise index and frequency argument
\begin{eqnarray}
\tilde{\boldsymbol{\gamma}}(\omega) &=\, \tilde{\boldsymbol{\gamma}}^\dagger(\omega) \,=& \tilde{\boldsymbol{\gamma}}^*(-\omega) \, , \label{eq:damp_dher} \\
\tilde{\boldsymbol{\nu}}(\omega) &=\, \tilde{\boldsymbol{\nu}}^\dagger(\omega) \,=& \tilde{\boldsymbol{\nu}}^*(-\omega) \, , \label{eq:noise_dher}
\end{eqnarray}
and by Bochner's theorem both $\tilde{\boldsymbol{\noisecorrelation}}(\omega)$ and $\tilde{\boldsymbol{\nu}}(\omega)$ are positive-definite for all frequencies.
Again the damping kernel $\tilde{\boldsymbol{\gamma}}(\omega)$ may be positive definite, negative definite, or indefinite.

\subsection{Classification of Damping Kernels} \label{sec:damping}
Environments with positive-definite damping kernels are \emph{damping} or \emph{resistive} environments,
while those with negative-definite damping kernels are \emph{amplifying}.
If the system coupling operators $\mathbf{L}_n$ are position operators, the damping terms correspond to forces linear in momentum.
Stationary correlations are the easiest to dissect and the most well behaved.
Their dissipation and damping kernels are related by
\begin{equation}
\tilde{\boldsymbol{\mu}}(\varepsilon) = \imath \varepsilon \, \tilde{\boldsymbol{\gamma}}(\varepsilon) \, ,
\end{equation}
and from the definition of the dissipation kernel in Eq.~\eqref{eq:muKernel} and the double Hermiticity in Eq.~\eqref{eq:damp_dher}-\eqref{eq:noise_dher},
the damping kernel will be most-generally positive or negative definite if we have a strict inequality between positive and negative energy argumented environmental correlations.
\begin{eqnarray}
\tilde{\boldsymbol{\noisecorrelation}}(-|\omega|) &>& \tilde{\boldsymbol{\noisecorrelation}}^*(+|\omega|)
\quad \mathrm{(Damping)}\, , \\
\tilde{\boldsymbol{\noisecorrelation}}(-|\omega|) &<& \tilde{\boldsymbol{\noisecorrelation}} ^*(+|\omega|)
\quad \mathrm{(Amplifying)}\, .
\end{eqnarray}
From Eq.~\eqref{eq:ideal_bath}, one can show that damping environments result when the initial stationary probability of the environment $p_\mathrm{E}(\varepsilon)$ is a monotonically decreasing function of the environment energy.
Amplifying environments result from monotonically increasing functions or \emph{population inversion}.
The most common example of each being positive and negative temperature reservoirs.

Given our damping representation of the master equation coefficients,
one can determine the dynamics of the system energy from the super-adjoint of the master equation \cite{Breuer02}:
\begin{equation}
\boldsymbol{\mathcal{L}}^\dagger \, \mathbf{H} = -\left\{ \dot{\mathbf{L}}_n , \left( \boldsymbol{\Gamma}_{nm} \diamond \dot{\mathbf{L}}_m \right) \right\} + \imath \left[  \dot{\mathbf{L}}_n , \left( \mathbf{N}_{nm} \diamond \mathbf{L}_m \right) \right] + \cdots \, ,
\end{equation}
where we have neglected any power generated by the slip and time-dependence intrinsic to the coupling.
Using the zeroth-order solution $\boldsymbol{\rho}(t) = \mathbf{G}_0(t) \, \boldsymbol{\rho}(0)$ and symmetries of the damping kernel,
the second-order expectation value for the cumulative energy lost through damping is given by
\begin{equation}
-\int_0^t \!\! d\tau_1 \int_0^t \!\! d\tau_2 \sum_{nm} \gamma_{nm}(\tau_1,\tau_2) \, \mathrm{Tr}\! \left[ \underline{\dot{\mathbf{L}}}_n(\tau_1) \, \boldsymbol{\rho}(0) \, \underline{\dot{\mathbf{L}}}_m(\tau_2) \right] \, ,
\end{equation}
which will be strictly dissipative for a positive-definite damping kernel.
This expression also contrasts \emph{nonlocal damping} to \emph{local damping}.
Delta correlated damping kernels are strictly dissipative at every instant of time whereas nonlocal damping kernels are only assured to be net dissipative in the full time accumulation.

\section{Non-Equilibrium Relations} \label{sec:NER}
\subsection{Non-Equilibrium Fluctuation-Dissipation Relation \& Inequality} \label{sec:FDR}
From the definitions of the multivariate noise kernel $\tilde{\boldsymbol{\nu}}(\omega)$ and damping kernel $ \tilde{\boldsymbol{\gamma}}(\omega)$, Eqs.~\eqref{eq:3kernels}-\eqref{eq:muKernel}, one can prove the \emph{fluctuation-dissipation inequality}:
\begin{equation}
\tilde{\boldsymbol{\nu}}(\omega) \geq \pm \, \omega \, \tilde{\boldsymbol{\gamma}}(\omega) \, , \label{eq:qnoise_bound}
\end{equation}
here for stationary correlations, and in the Fourier domain where the $\omega$ would denote energy level transitions of the system.
To prove this one simply notes that the noise kernel is the sum of two positive-definite kernels whereas the dissipation kernel is given by their difference.
The essential point is that if there is any damping, or amplification, there will be quantum noise and Eq.~\eqref{eq:qnoise_bound} determines its lower bound.
This is quite a departure from classical physics where noise can be made to vanish in the zero-temperature limit,
although the lower bound of this inequality is satisfied by zero-temperature quantum noise since $\tilde{\boldsymbol{\noisecorrelation}}(|\omega|)=0$ in that case.

For the case of one collective system coupling, coupled to one or more environments, it is sufficient to define a fluctuation-dissipation relation
\begin{equation}
\tilde{\nu}(\omega) = \tilde{\kappa}(\omega) \, \tilde{\gamma}(\omega) \, , \\
\tilde{\kappa}(\omega) \equiv \frac{\tilde{\nu}(\omega)}{\tilde{\gamma}(\omega)} \, ,
\end{equation}
with $\tilde{\kappa}(\omega)$ the fluctuation-dissipation kernel \cite{HPZ92,HPZ93} which relates fluctuations to dissipation.
For multivariate noise one might use the symmetrized product
\begin{equation}
\tilde{\boldsymbol{\nu}}(\omega) = \frac{1}{2} \left[ \tilde{\boldsymbol{\kappa}}(\omega) \, \tilde{\boldsymbol{\gamma}}(\omega) + \tilde{\boldsymbol{\gamma}}(\omega) \, \tilde{\boldsymbol{\kappa}}(\omega) \right] \, , \label{eq:multiFDR}
\end{equation}
which would ensure $\tilde{\boldsymbol{\kappa}}(\omega)$ to be positive definite if $\tilde{\boldsymbol{\gamma}}(\omega)$ is,
in accord with this being a (continuous) Lyapunov equation \cite{Bhatia07}.
We will use this particular definition for quantum Brownian motion in the next section.
Inequality \eqref{eq:qnoise_bound} can now be restated as
\begin{equation}
\tilde{\boldsymbol{\kappa}}(\omega) \geq | \omega | \, ,
\end{equation}
for damping environments.
Typically $\tilde{\boldsymbol{\kappa}}(\omega)$ will contain dependence upon the precise nature of the environment couplings $\mathbf{l}_n$.

\subsection{Equilibrium Fluctuation-Dissipation Relation}
Let us consider a constant system-environment interaction, constant environment Hamiltonian, and initial stationary probabilities of the environment given by $p_\mathrm{E}(\varepsilon)$.
If the FDR is to be independent of precisely how the system and environment are coupled,
then one can work out from the microscopic theory that the FDR kernel must be a scalar quantity, directly related to the initial state of the environment by way of
\begin{equation}
\frac{\tilde{\kappa}(\omega)}{\omega} = \frac{p_\mathrm{E}(\varepsilon - \omega) + p_\mathrm{E}(\varepsilon)}{p_\mathrm{E}(\varepsilon - \omega) - p_\mathrm{E}(\varepsilon)} \, ,
\end{equation}
for all $\varepsilon$.
To prove this one first applies relation \eqref{eq:ideal_bath} to definitions \eqref{eq:nuKernel}-\eqref{eq:muKernel},
and notes that if the dissipation and noise are related in a manner independent of the coupling
then the two kernels must be related term-by-term in a sum over couplings.

But such an equality between the FDR kernel and mode probabilities implies the functional relation
\begin{equation}
p_\mathrm{E}(\varepsilon - \omega) = \left[\frac{\frac{\tilde{\kappa}(\omega)}{\omega}+1}{\frac{\tilde{\kappa}(\omega)}{\omega}-1}\right] p_\mathrm{E}(\varepsilon) \, ,
\end{equation}
where the $\omega$ translations can factor out.
This factorization property is unique to exponential functions;
therefore, only the thermal distribution $p_\mathrm{E}(\varepsilon) \propto e^{-\beta \varepsilon}$ can produce a fluctuation-dissipation relation which is generally coupling independent.
We then have that
\begin{equation}
\tilde{\kappa}_T(\omega) \equiv \omega \coth\!\left( \frac{\omega}{2T} \right) \, , \label{eq:KappaT}
\end{equation}
for the thermal distribution.
One should be careful to note that the thermal FDR is not special because it exists, nor because of its simple form,
but because of its invariance to couplings (model-independence).
In a more general context, the thermal FDR is also special because it ensures a relaxation to detailed balance in a coupling-invariant manner.
In fact, these properties can be shown to be equivalent \cite{QOS}.

As a concrete example of an elegant yet non-equilibrium FDR,
the late-time dominating stationary correlations for linear coupling to a squeezed thermal reservoir \cite{HuMatacz94,Koks97} will produce the FDR kernel
\begin{equation}
\tilde{\kappa}_T^{\,r}(\omega) = \cosh\!\left[ 2\, r(\omega) \right] \omega \coth\!\left( \frac{\omega}{2T} \right) \, ,
\end{equation}
where $r(\omega)$ is the squeezing parameter, which may be allowed to vary with the energy scale.
One can easily see that this FDR also obeys inequality \eqref{eq:qnoise_bound} as it must.

\section{Non-Equilibrium Uncertainty Principle} \label{sec:QBM}
In the context of second-order perturbation theory, quantum noise is effectively Gaussian in the influence functional, and Gaussian noise is equivalent to that arising from linear coupling to a bath of harmonic oscillators.
Therefore any violations of Eq.~\eqref{eq:qnoise_bound} must correspond to environment oscillators in a non-quantum state.
In the phase-space or Wigner function representation \cite{Hillery84}, HUP violating states of the environment can be constructed which violate the quantum FDI.
Such is the case for the classical vacuum, which has vanishing fluctuations yet finite damping.
Now we shall show that FDI violating noise can also relax the system into a HUP violating state.

Let us consider weakly influencing a system of oscillators at resonance, all with mass $m$ and frequency $\omega$,
via position-position coupling to some phenomenological set of noise processes with resistive correlation $\tilde{\boldsymbol{\noisecorrelation}}(\omega)$.
We do not assume the system-environment couplings to be identical, nor do we neglect the presence of cross-correlations among the noise processes.

From the results of Ref.~\cite{QBM,NQBM} and the second-order master equation coefficients \eqref{eq:WCGME},
the damping kernel $\tilde{\boldsymbol\gamma}(\omega)$ will play the role of the dissipation coefficients
and the noise kernel $\tilde{\boldsymbol\nu}(\omega)$ will play the role of the normal diffusion coefficients in the Fokker-Planck or master equation.
Integrals over the two kernels will then provide the system renormalization and anti-diffusion coefficients respectively.
Given sufficient dissipation and bandwidth-limited correlations, the system will relax into a Gaussian state which satisfies the Lyapunov equation
\begin{equation}
\tilde{\boldsymbol\nu}(\omega)  = \frac{1}{2} \left[ \left(\frac{2}{m} \boldsymbol{\sigma}_{\!pp}\right) \tilde{\boldsymbol\gamma}(\omega) + \tilde{\boldsymbol\gamma}(\omega) \left(\frac{2}{m} \boldsymbol{\sigma}_{\!pp}\right) \right] \, , \label{eq:LyapunovQBM}
\end{equation}
for the momentum covariance,
which has elements $\left\langle p_i  p_j \right\rangle$,
and
\begin{eqnarray}
\boldsymbol{\sigma}_{\!xx} &=& \frac{1}{(m\omega)^2} \boldsymbol{\sigma}_{\!pp} \, , \\
\boldsymbol{\sigma}_{\!xp} &=& \mathbf{0} \, ,
\end{eqnarray}
for the remaining covariances in the phase-space (Wigner function) representation \cite{Hillery84},
and to lowest order in the system-environment interaction.
Comparing Eq.~\eqref{eq:LyapunovQBM} to Eq.~\eqref{eq:multiFDR}, we can express our covariances
\begin{eqnarray}
\boldsymbol{\sigma}_{\!xx} &=& \frac{1}{2m\omega^2} \tilde{\boldsymbol\kappa}(\omega) \, , \\
\boldsymbol{\sigma}_{\!pp} &=& \frac{m}{2} \tilde{\boldsymbol\kappa}(\omega) \, ,
\end{eqnarray}
in terms of the FDR kernel $\tilde{\boldsymbol\kappa}(\omega)$.
So far our FDR kernel remains phenomenological and not microscopically derived,
however it must be positive definite for this to describe a physical state.
As $\tilde{\boldsymbol\kappa}(\omega)$ is positive definite, we may transform to the basis in which it is diagonalized.
If $\tilde{\boldsymbol\kappa}(\omega)$ is a scalar quantity, then this is simply the normal basis of the free system.
For each mode in this basis we have the phase-space covariance
\begin{equation}
\boldsymbol{\sigma}_{\!n} = \left[ \begin{array}{cc} {\sigma}_{\!n}^{xx} & {\sigma}_{\!n}^{xp} \\ {\sigma}_{\!n}^{px} & {\sigma}_{\!n}^{pp} \end{array} \right] \,=\,  \left[ \begin{array}{cc}  \frac{1}{2m\omega^2} \tilde{\kappa}_n(\omega)  & 0 \\ 0 & \frac{m}{2} \tilde{\kappa}_n(\omega) \end{array} \right] \, ,
\end{equation}
which must then satisfy the generalized Heisenberg uncertainty relation due to Schr\"{o}dinger \cite{Robertson34,Trifonov02}:
\begin{equation}
\langle \Delta x^2 \rangle \langle \Delta p^2 \rangle - \frac{1}{2} \left\langle \{ \Delta x , \Delta p \} \right\rangle \geq \frac{1}{4} \, ,
\end{equation}
or in terms of the phase-space covariance
\begin{equation}
\mathrm{det}( \boldsymbol{\sigma} ) \geq \frac{1}{4} \,
\end{equation}
and, therefore, it must be the case that
\begin{equation}
\tilde{\kappa}_{n}(\omega) \geq \omega \, ,
\end{equation}
for all $\omega > 0$.
But this is equivalent to our previous statement
\begin{equation}
\tilde{\boldsymbol\kappa}(\omega) \geq \omega \, ,
\end{equation}
in terms of positive definiteness as $\omega$ is a scalar quantity.
So not only do FDI violating correlations arise from HUP violating states, they can also produce HUP violating states via dissipation and diffusion (and decoherence).

Furthermore we can say that in the weak-damping limit,
the scalar FDR kernel $\tilde{\kappa}(\omega)$ precisely determines the (asymptotic) non-equilibrium uncertainty product 
\begin{equation}
\mathrm{det}( \boldsymbol{\sigma} ) = \left( \frac{1}{2} \frac{\tilde{\kappa}(\omega)}{\omega} \right)^2 \, \label{eq:Uprod},
\end{equation}
for a single system mode of frequency $\omega$.
Larger FDR kernels naturally produce larger uncertainty and minimal FDR kernels (zero temperature) produce minimal uncertainty.
Non-perturbative results require evaluation of the exact expressions found in Refs.~\cite{QBM,NQBM} for a single system oscillator and multiple system oscillators respectively.

\section{Discussion}

In this paper we have derived a fluctuation-dissipation inequality (FDI) for an open quantum system interacting with a non-equilibrium environment from the microscopically-derived environment correlation function and recovered the well-known fluctuation-dissipation relation (FDR) for a thermal environment.
The FDI is a very general statement contrasting quantum noise to classical noise, and is satisfied even for non-equilibrium fluctuations.
Simply put, quantum fluctuations are bounded below by quantum dissipation, whereas classically the fluctuations can be made to vanish.
The lower bound of this inequality is exactly satisfied by zero-temperature noise and is in accord with the Heisenberg uncertainty principle (HUP).
FDI violating correlations arise from HUP violating states of the environment and can relax the open system into HUP violating states.
Therefore, the FDI can be viewed as an open-system corollary to the HUP both from microscopic and phenomenological considerations.
Analogously, the non-equilibrium FDR also determines the non-equilibrium uncertainty product, 
most directly in the limit of weak-damping [cf.\ Eq.~(\ref{eq:Uprod})], and the corresponding FDI implies the HUP.

\acknowledgments

This work is supported in part by NSF grants PHY-0426696, PHY-0801368, DARPA grant DARPAHR0011-09-1-0008 and the Laboratory for Physical Sciences.

\bibliography{bib}{}
\bibliographystyle{apsrev}
\end{document}